\documentclass[9pt,twocolumn,twoside]{pnas-new}
\templatetype{pnasresearcharticle}
\usepackage{graphicx}%
\usepackage[font=small]{caption}
\usepackage[labelformat=empty, position=top]{subcaption}
\usepackage[export]{adjustbox}
\usepackage{multirow}%
\usepackage{amsmath,amssymb,amsfonts}%
\usepackage{amsthm}%
\usepackage{mathrsfs}%
\usepackage[title]{appendix}%
\usepackage{xcolor}%
\usepackage{textcomp}%
\usepackage{manyfoot}%
\usepackage{booktabs}%
\usepackage{algorithm}%
\usepackage{algorithmicx}%
\usepackage{algpseudocode}%
\usepackage{listings}%
\theoremstyle{thmstyleone}%
\theoremstyle{thmstyletwo}%
\theoremstyle{thmstylethree}%
\usepackage{multicol}
\usepackage{soul}
\soulregister{\cite}{7}
\usepackage{xcolor}
\usepackage{colortbl}

\begin{document}

\title{A generalizable framework for unlocking missing reactions in genome-scale metabolic networks using deep learning}

\author[a]{Xiaoyi Liu}
\author[a]{Hongpeng Yang}
\author[b]{Chengwei Ai}
\author[c]{Ruihan Dong}
\author[d]{Yijie Ding}
\author[e,f,1]{Qianqian Yuan}
\author[g,h,1]{Jijun Tang}
\author[b,1]{Fei Guo}

\affil[a]{University of South Carolina, Columbia, 29208, USA}
\affil[b]{Central South University, Changsha, 410083, China}
\affil[c]{Peking University, Beijing, 100871, China}
\affil[d]{Yangtze Delta Region Institute, University of Electronic Science and Technology of China, Quzhou, 324000, China}
\affil[e]{Biodesign Center, Key Laboratory of Engineering Biology for Lowcarbon Manufacturing, Tianjin Institute of Industrial Biotechnology, Chinese Academy of Sciences, Tianjin, 300308, China}
\affil[f]{National Technology Innovation Center of Synthetic Biology, Tianjin, 300308, China}
\affil[g]{Shenzhen Institute of Advanced Technology, Chinese Academy of Sciences, Nanshan, 518055, China}

\leadauthor{Liu et al.}

\significancestatement{Genome-scale Metabolic Models (GEMs) provide a comprehensive understanding of gene-reaction-metabolite interactions and are widely used as a powerful tool in bioengineering applications, particularly in microbial product manufacturing. However, even the most highly-curated GEMs remain gaps due to limited knowledge of metabolic processes. Traditional methods are resource-intensive and restricted to known pathways. To address this issue, we introduce CLOSEgaps, a model-free, data-driven deep learning framework that integrates hypergraph Convolutional Network and attention mechanism to predict metabolic gaps. By combining the robust CLOSEgaps framework with a simulation workflow, we have automated the gap-filling process, offering the potential to expedite the completion of GEMs and facilitate more effective bioengineering efforts.}

\authorcontributions{F.G. conceived and designed the project. X.L. constructed the database and developed the algorithm. All authors analyzed the results. X.L. prepared the manuscript. F.G. edited and approved the manuscript.}
\authordeclaration{The authors declare no competing financial interests.}
\correspondingauthor{\textsuperscript{1}To whom correspondence should be addressed. E-mail: guofei@csu.edu.cn, yuan\_qq@tib.cas.cn or jj.tang@siat.ac.cn}

\keywords{genome-scale metabolic models $|$ gap-filling $|$ missing annotation $|$ hypergraph learning $|$ hyperlink prediction}

\begin{abstract}
Incomplete knowledge of metabolic processes hinders the accuracy of GEnome-scale Metabolic models (GEMs), which in turn impedes advancements in systems biology and metabolic engineering. Existing gap-filling methods typically rely on phenotypic data to minimize the disparity between computational predictions and experimental results. However, there is still a lack of an automatic and precise gap-filling method for initial state GEMs before experimental data and annotated genomes become available. In this study, we introduce CLOSEgaps, a deep learning-driven tool that addresses the gap-filling issue by modeling it as a hyperedge prediction problem within GEMs. Specifically, CLOSEgaps maps metabolic networks as hypergraphs and learns their hyper-topology features to identify missing reactions and gaps by leveraging hypothetical reactions. This innovative approach allows for the characterization and curation of both known and hypothetical reactions within metabolic networks. Extensive results demonstrate that CLOSEgaps accurately gap-filling over $96\%$ of artificially introduced gaps for various GEMs. Furthermore, CLOSEgaps enhances phenotypic predictions for $24$ GEMs and also finds a notable improvement in producing four crucial metabolites (Lactate, Ethanol, Propionate, and Succinate) in two organisms. As a broadly applicable solution for any GEM, CLOSEgaps represents a promising model to automate the gap-filling process and uncover missing connections between reactions and observed metabolic phenotypes.
\end{abstract}

\dates{This manuscript was compiled on \today}
\doi{\url{www.pnas.org/cgi/doi/10.1073/pnas.XXXXXXXXXX}}

\maketitle
\thispagestyle{firststyle}
\ifthenelse{\boolean{shortarticle}}{\ifthenelse{\boolean{singlecolumn}}{\abscontentformatted}{\abscontent}}{}

\firstpage[8]{2}

\dropcap{I}ntegrating a comprehensive understanding of biology at the systems level is essential for advancing bioengineering, drug targeting, and medical therapies \cite{robinson2017anticancer, kim2021machine, xu2022novo}. In this pursuit, metabolic networks and annotated genomes are leveraged to gain a holistic view of cellular functions. Despite these efforts, gaps still exist in our knowledge of cellular metabolic capabilities. Thus, systematically uncovering these unknown metabolic processes has the potential to catalyze a wide range of medical and biotechnological applications \cite{vayena2022workflow}. As a mathematical representation of an organism's metabolism, GEnome-scale Metabolic models (GEMs) offer comprehensive gene-reaction-metabolite connectivity through stoichiometric and reaction-gene matrices. These models have emerged as a powerful tool for systematically analyzing cellular metabolic functions \cite{thiele2014fastgapfill, o2015using, gu2019current, li2021bayesian, li2022improving, liu2023mvml, mao2023cave}. With extensive use in the study of model organisms, these models are commonly evaluated through simulation techniques such as Flux Balance Analysis (FBA) \cite{orth2010flux}, which assumes a balanced flux of metabolites in the metabolic network via linear optimization \cite{gu2019current, domenzain2022reconstruction}. Recently, the availability of whole-genome sequencing data \cite{nayfach2021genomic} and automatic reconstruction pipelines \cite{machado2018fast, zimmermann2021gapseq} have opened up new avenues for constructing draft GEMs. However, incomplete knowledge of metabolic processes and incomplete genomic and functional annotations results in incomplete draft GEMs, characterized by missing reactions \cite{thiele2014fastgapfill, vayena2022workflow, zimmermann2021gapseq}. This presents an opportunity for completing GEMs through the gap-filling process, aimed at minimizing the number of missing reactions by adding reactions to the model \cite{thiele2010protocol, pan2018advances}. Hence, effective and robust gap-filling algorithms are essential for metabolic network reconstructions  \cite{oyetunde2017boostgapfill, latendresse2012construction}.

Recently, various classic gap-filling algorithms have been developed and reviewed, including constraint-based modeling, GrowMatch, and comparative genomics methods \cite{orth2010systematizing}. However, these traditional gap-filling methods rely on phenotypic data and extensive manual curation to address knowledge gaps in draft GEMs, impacting the time, accuracy, and effectiveness of the GEM models in biomedical applications \cite{orth2010systematizing, pan2018advances, rana2020recent, bernstein2021addressing}. These methods are also limited by their reliance on experimental data, which is often unavailable for non-model or “uncultivable” organisms, and even when available, making the process costly and time-consuming to obtain through phenotypic screening \cite{chen2023teasing, orth2010systematizing, vayena2022workflow}. To address these limitations, successful gap-filling needs a more practical approach. Thus, topology-based approaches have gained popularity in the field of bioinformatics for link prediction \cite{grover2016node2vec, schlichtkrull2018modeling, hamilton2017inductive}. For Neural Hyperlink Predictor (NHP) \cite{yadati2020nhp} and Coordinated Matrix Minimization (CMM) \cite{zhang2018beyond} are two hypergraph-based methods that can be used to efficiently gap-filled the artificially introduced gaps of GEMs. However, these methods are limited to the space of known annotated proteins and biochemistry, Expanding our understanding of metabolic networks to include novel biochemistry requires exploration beyond the space of known biochemistry \cite{vayena2022workflow}, also a crucial step in advancing our understanding of metabolic network function.

Herein, we introduce a novel framework named CLOSEgaps (hypergraph ConvoLution netwOrk and attention mechaniSm integrated Explorer for GAPS prediction of metabolism (Fig.~\ref{fig1})) to complete highly incomplete GEMs at the reaction level. CLOSEgaps addresses the limitations of current gap-filling methods by using a deep learning-based approach that predicts missing reactions in GEMs solely through the topological features of metabolic networks, without relying on experimental data, thereby overcoming the dependency on incomplete or unavailable biochemical annotations. The extensive experimental results demonstrate that this hypergraph-based strategy significantly improves gap-filling performance, with accuracy reaching over $96\%$. The diverse and multi-modal nature of deep learning-based CLOSEgaps through heatmap visualized and reflects the actual metabolic relations in GEMs. 

To further assess the improvement of the gap-filling process, CLOSEgaps was applied to $24$ draft GEMs reconstructed from the wildly used pipeline, CarveMe \cite{machado2018fast}. By integrating GEMs with the pool of hypothetical reactions (from the universal BiGG reaction pool), we aim to identify missing metabolic reactions predicted by CLOSEgaps and bridge the gap in understanding metabolic network function. We show that CLOSEgaps suggests novel biochemistry and significantly improves the theoretical predictability of metabolic fermentation metabolites. The combination of the robust CLOSEgaps framework, hypothetical reactions database, and simulation methods fully automate the gap-filling process, offering the potential to accelerate the completion of GEMs and enable effective bioengineering.

\section{Results}

\subsection*{A Workflow of CLOSEgaps for Metabolic Network Reconstruction} \label{sec3}

\begin{figure*}
    \centering
    \includegraphics[width=\linewidth]{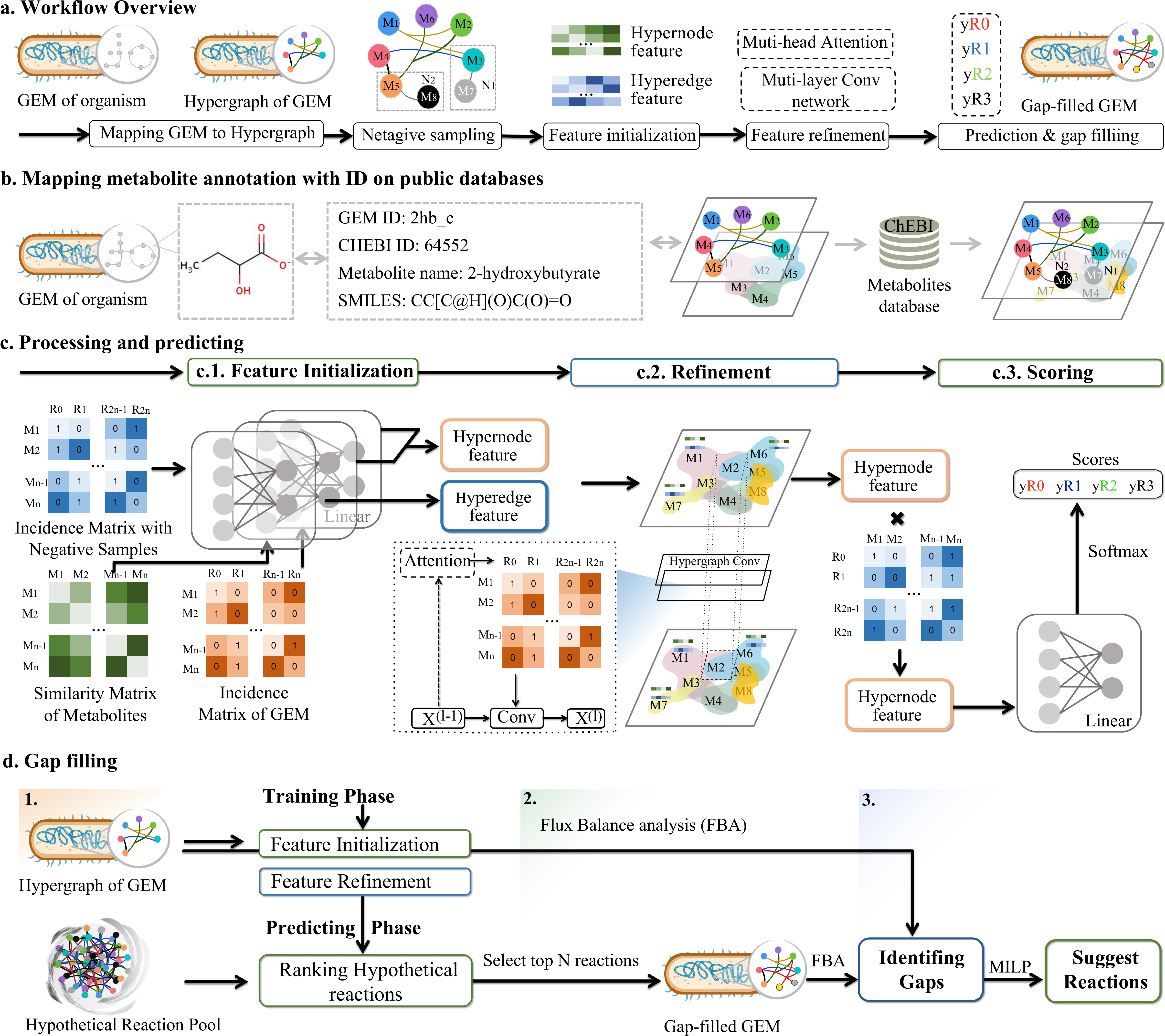}
    \caption{\textbf{a} The pipeline of CLOSEgaps. We formulated the process as five distinct phases: (1) mapping GEM to hypergraph, (2) negative sampling, (3) feature initialization, (4) feature refinement, and (5) prediction or gap-filling. \textbf{b} Mapping GEM to hypergraph with BiGG reactions and ChEBI metabolites database, metabolites are represented by SMILES. The processed ChEBI database is used for negative sampling. \textbf{c} Processing and predicting. \textbf{c.1} The incidence matrix of the hypergraph (incidence matrix of GEM and incidence matrix with negative samples), and the similarity of the metabolites matrix are used to initialize features through a fully connected layer. \textbf{c.2} The hypergraph convolution and hypergraph attention networks are used to refine hypernode and hyperedge features. \textbf{c.3} The ranking module predicts missing reactions. \textbf{d} The gap-filling inference workflow: (\textbf{1}) The draft GEM and hypothetical database as input, GEM are fully used as the training set and ranking each reaction in hypothetical reaction pool, (\textbf{2}) FBA is utilized to predict fermentation phenotypes for the gap-filled GEMs and the wild-type GEMs, and (\textbf{3}) MILP causally suggests the missing reactions for the production of phenotypes.}
    \label{fig1}
\end{figure*}

CLOSEgaps is used in the workflow to predict missing reactions in GEMs. As shown in Fig.~\ref{fig1}a, it involves five key steps: mapping GEM to hypergraph, negative reaction sampling, feature initialization, feature refinement, and prediction or ranking. In the first step (Fig.~\ref{fig1}b), we mathematically map the GEM to an unweighted hypergraph (see “\textit{Material and Methods}” and “\textit{Data Collection and Preprocessing} in \textit{SI Appendix}”) The second step uses a metabolic network and the ChEBI database \cite{degtyarenko2007chebi} to sample negative reactions. The third step maps metabolites to hypernode features and reactions to hyperedge features and applies a fully connected layer for feature initialization (Fig.~\ref{fig1}c.1). The fourth step (Fig.~\ref{fig1}c.2) refines the metabolic network structure and properties with hypergraph convolution and attention. In the final step (Fig.~\ref{fig1}c.3 and d), the hyperedge feature is updated and multiplied by the transposed incidence matrix, and then each reaction's feature vector is fed into a neural network to determine its confidence level.

\subsection*{Metabolic Network Reconstruction}\label{sec4}

\subsubsection*{Assessment of Robustness on Artificially Introduced Gaps}\label{sec10}

The automated reconstruction of GEMs hinges upon the accuracy of the prediction model for missing reactions. Therefore, it is imperative to attain a dependable prediction of missing reactions for GEMs. The goal of this validation is to test the ability of CLOSEgaps to recover artificially introduced gaps (i.e., artificially removing existing reactions in metabolic networks). Thus, to examine the generalizability performance of CLOSEgaps, we tested CLOESgaps with CHESHIRE \cite{chen2023teasing}, GraphSAGE \cite{hamilton2017inductive}, NHP \cite{yadati2020nhp}, GCN and Node2Vector \cite{kipf2016semi,grover2016node2vec}, RGNN \cite{mitra2024knowledge,schlichtkrull2018modeling} and HGNN \cite{feng2019hypergraph} on five high-quality BiGG GEMs and two organic chemistry datasets as summarized in Table~\ref{dataset_statistics}.

Below we conducted two types of validation based on artificially introduced gaps. For both types, CLOSEgaps requires a complete training process for each specific GEM it is applied to, ensuring that the model is fully tailored to the specific metabolic network being analyzed. For each GEM, negative (fake) reactions were sampled at $1$:$1$ ratio to all positive reactions, by replacing $50\%$ of the metabolites in each reaction with randomly selected metabolites from the CHEBI database (see “\textit{Hypergraph and negative sampling} in \textit{Materials and Methods}”). Accordingly, after combining the positive and negative reactions for each GEM, we randomly selected $60\%$ of the data for the training, $20\%$ for validation, and $20\%$ for testing, with the testing set treated as missing reactions. This process was repeated over $10$ independent Monte Carlo runs (see “\textit{CLOSEgaps Model} in \textit{Materials and Methods}”). Furthermore, we conducted extensive experiments with various negative sampling strategies (detailed results available in \textit{SI Appendix}), such as $50\%$, $20\%$, and $80\%$ metabolite replacement, and spanning $1$:$1$, $1$:$2$, and $1$:$3$ negative-to-positive reaction ratios (see \textit{SI Appendix}, Fig.~S2). And an optimal atom-balanced negative reaction sampling strategy, by preserving the atomic count consistency between reactants and products (see \textit{SI Appendix}, Table~S2-S3). 

\begin{table*}[htpb]\centering
    \caption{Metabolic network and chemical reaction dataset statistics.}
    \label{dataset_statistics}{
    \begin{tabular}{lcccc}
        \toprule Dataset & Species & Metabolites (vertices) & Reactions (hyperlinks)\\
        \midrule
        Yeast8.5&\textit{Saccharomyces cerevisiae} (Jul. 2021) & 1136 & 2514\\
        iMM904&\textit{Saccharomyces cerevisiae} S288C (Oct. 2019) & 533 & 1026\\
        iAF1260b&\textit{Escherichia coli} str. K-12 substr. MG1655 & 765 & 1612\\
        iJO1366&\textit{Escherichia coli} str. K-12 substr. MG1655 & 812 & 1713\\
        iAF692&\textit{Methanosarcina barkeri} str. \textit{Fusaro}&422&562\\
        USPTO\_$3k$&Chemical reaction&6706&3000\\
        USPTO\_$8k$&Chemical reaction&15405&8000\\
        \bottomrule
    \end{tabular}
    }
\end{table*}

To perform the first type validation, we tested CLOSEgaps on five high-quality BiGG GEMs with $60\%$ training and $40\%$ validation and testing. Following the previous missing reaction prediction works \cite{zhang2018beyond, yadati2020nhp, chen2023teasing}, CLOSEgaps achieves the best performance in four classical classification performance metrics: the F1 score, the area under the receiver operating characteristic curve (AUC), the area under the precision-recall (AUPR), and precision and recall. A threshold score of $0.5$ was used to determine whether a test reaction is true or false.

\begin{figure*}
    \centering
    \includegraphics[width=\linewidth]{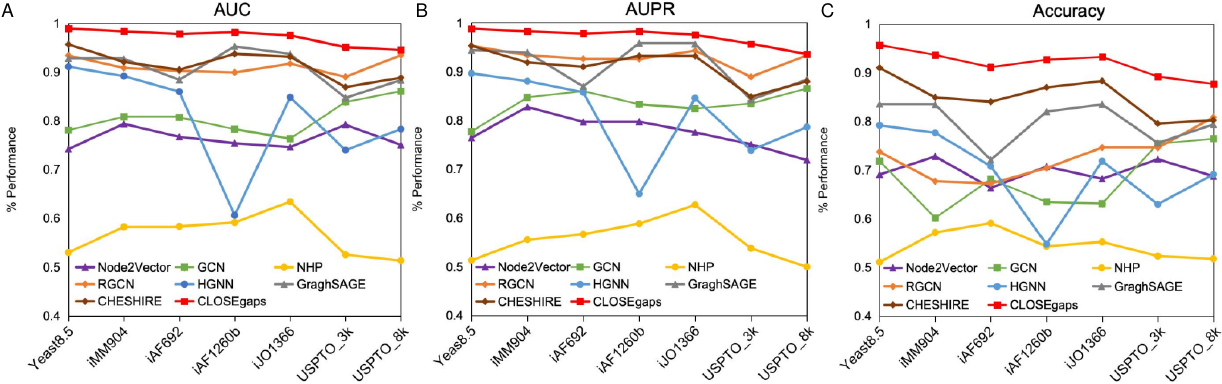}
    \caption{\textbf{Performance validation using artificially introduced gaps.} \textbf{(A}, \textbf{B}, \textbf{C)} Boxplots of the performance metrics (AUC, AUPR, Accuracy) calculated on 7 datasets (each dot represents a dataset) for CLOSEgaps vs.CHESHIRE, GraphSAGE, HGNN, RGNN, NHP, GCN, and Node2Vector.}
    \label{AUC_AUPR}
\end{figure*}

\begin{figure*}
    \centering
    \includegraphics[width=17.5cm]{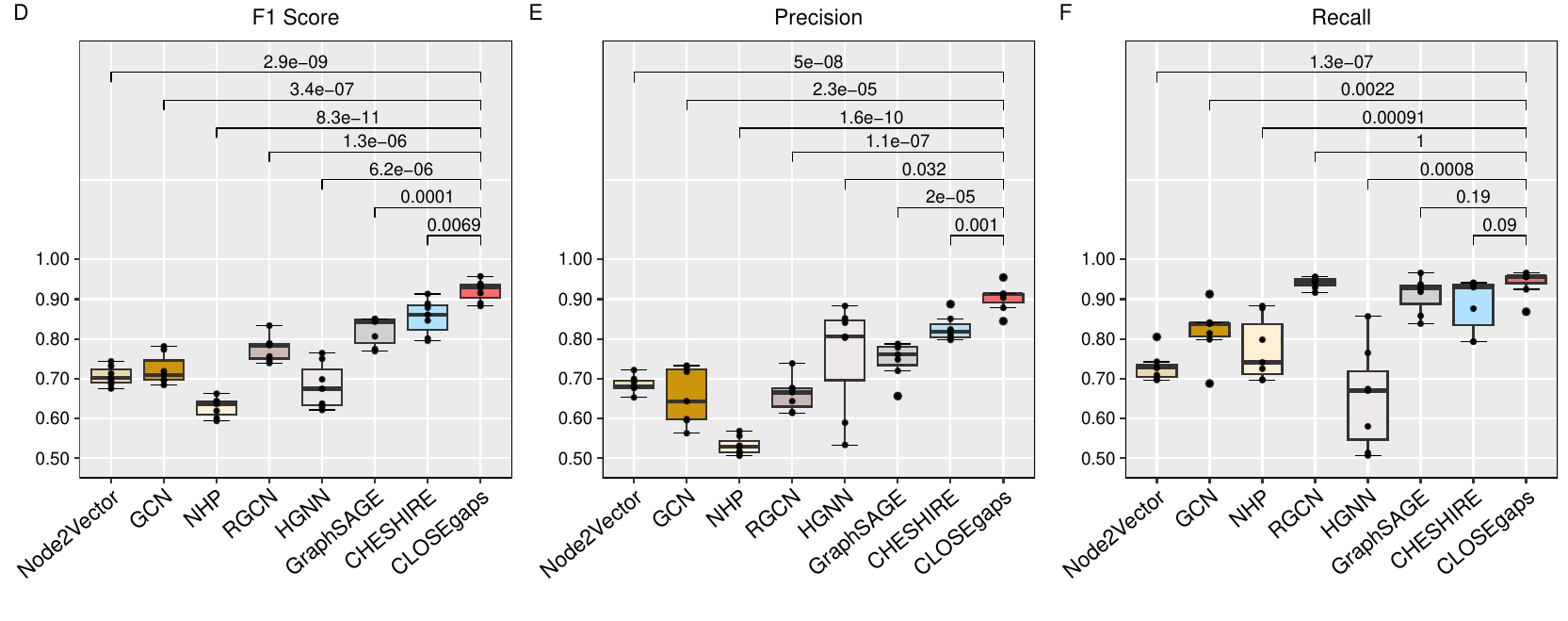}
    \caption{\textbf{Performance validation using artificially introduced gaps.} \textbf{{(A}, \textbf{B}, \textbf{C)}} Boxplots of the performance metrics (F1 score, Precision, and Recall) calculated on 7 datasets (each dot represents a dataset) for CLOSEgaps vs.CHESHIRE, GraphSAGE, HGNN, RGNN, NHP, GCN, and Node2Vector.}
    \label{boxplot}
\end{figure*}

As illustrated in Fig.~\ref{AUC_AUPR} and Fig.~\ref{boxplot}, CLOSEgaps exhibited superior performance, achieving the highest average over $97\%$ AUC and AUPR. A \textit{Saccharomyces cerevisiae} yeast8.5 metabolic network dataset from Lu et al. \cite{lu2019consensus} was used for evaluation. CLOSEgaps attains an F1 score of $96\%$, AUC of $99\%$, AUPR of $99\%$, Precision of $95\%$, and Recall of $96\%$. CLOSEgaps outperformed Node2Vector \cite{grover2016node2vec} by approximately $25\%$ and exceeded other graph embedding approaches by nearly $20\%$ across these five evaluation metrics, demonstrating the advantages of our proposed hypergraph framework.

Additionally, with model iMM904, where CLOSEgaps is able to achieve AUC $=98.38\%$, AUPR $=98.28\%$, Recall $=96.6\%$, Precision $=91.28\%$, and F1 score $=93.87\%$ on the testing set, demonstrating its accuracy in predicting missing reactions with any \textit{S.cerevisiae} GEMs. Moreover, CLOSEgaps also demonstrated its generalized capability with an AUC and AUPR of $98.21\%$, $98.27\%$ and $97.55\%$, $97.58\%$ on the iAF1260b and iJO1366 \textit{E.coli} models, respectively. Additionally, CLOSEgaps achieved a $97.84\%$ AUC and $97.83\%$ AUPR for annotating missing reactions with the iAF692 \textit{Methanosarcina barkeri} model. Notably, CLOSEgaps consistently outperformed topology-based models such as CHESHIRE, GraphSAGE, HGNN, RGNN, NHP, GCN, and Node2Vector, with increases of up to $7.35\%$ and $6.84\%$ in AUC and AUPR, respectively. 

The second validation considers the universality of CLOSEgaps and the complexity of organic reactions. We used the same unbalanced atom number strategy with negative reactions at a $1:1$ ratio to positive reactions with $60\%$ training and $40\%$ validation and testing. This experiment highlights that CLOSEgaps can predict reactions not only within biological datasets but also exhibits adaptability in learning and discerning chemical reaction datasets. Notably, the organic reactions experiment does not correlate with the GEM missing reaction prediction task.

The results demonstrated that CLOSEgaps attained the highest AUC of $95.13\%$, AUPR of $95.72\%$ on the United States Patent and Trademark Office (USPTO) dataset \cite{lowe2012extraction} with $3,000$ reactions called USPTO\_$3k$, and $94.56\%$ and $93.62\%$ on the USPTO dataset with $8,000$ reactions called USPTO\_$8k$), signifying its reliability and accuracy in predicting missing reactions not only in biological networks but also in various chemical reaction networks (as shown in Fig.~\ref{boxplot}). Nevertheless, it is essential to note that the chemical reaction data explored in this study does not pertain to the application of metabolic network reconstruction.

\subsubsection*{Assessment of robustness on highly incomplete GEMs}\label{sec12}

To assess the robustness of CLOSEgaps in gap-filling GEMs with significant incompleteness during the initial reconstruction, we carried out experiments on four GEMs from three species, namely yeast8.5, iMM904, iAF692, and iAF1260b. Rather than using a fixed cutting score, we employed a number of recovered reactions as a second measurement \cite{zhang2018beyond}. Specifically, if $N$ reactions are missing, we measure how many of the top $N$ predictions are correct. In our study, $20\%$, $40\%$, $60\%$, and $80\%$ of metabolic reactions were randomly removed to introduce hypothetical gaps in each of the GEMs. The remaining reactions were utilized as the training set, while those that were removed served as the testing set.

\begin{figure*}[htpb]
    \centering
    \includegraphics[width=15.5cm]{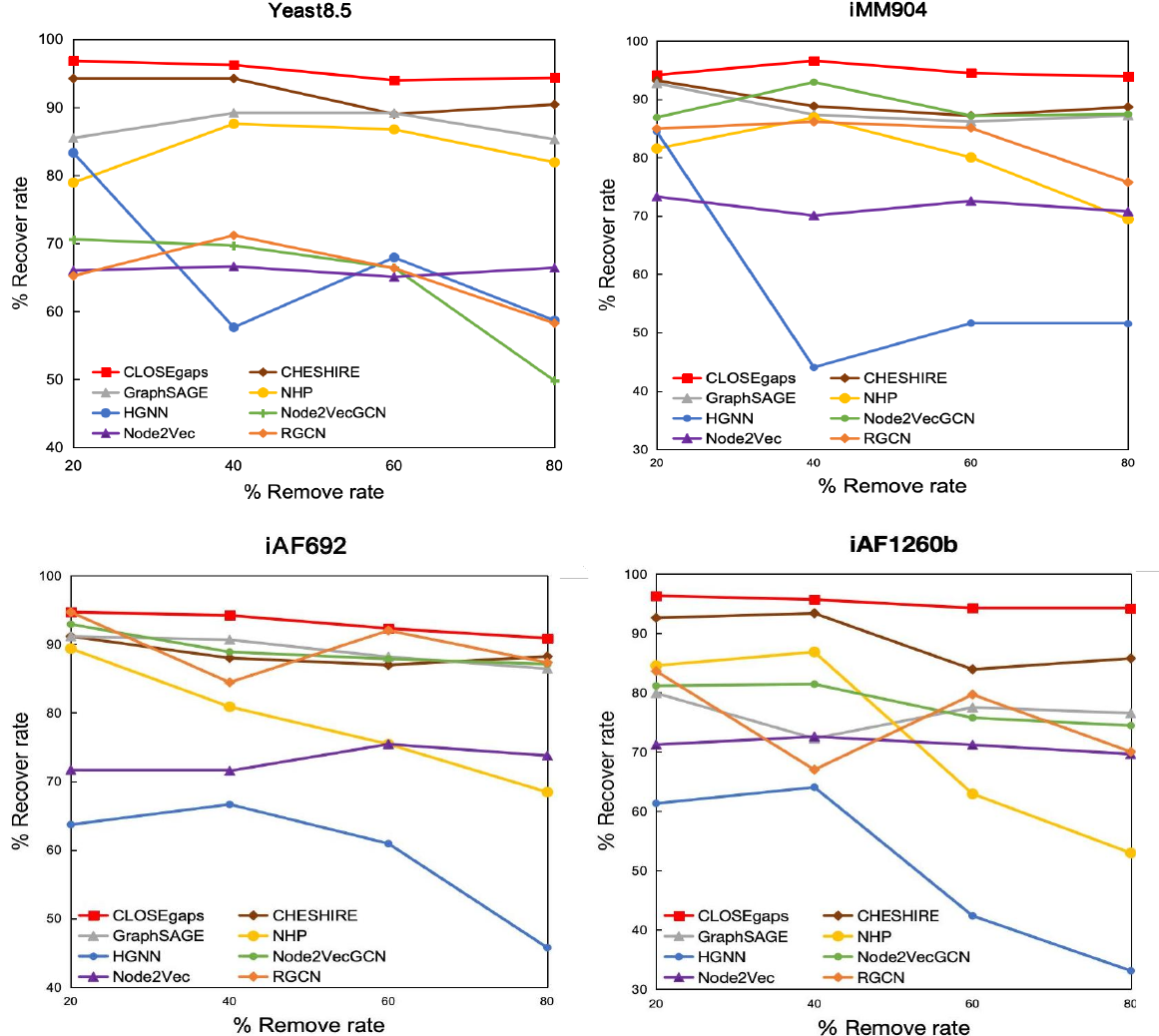}
	\caption{Comparison of CLOSEgaps with other methods (CHESHIRE, GraphSAGE, HGNN, RGNN, NHP, GCN, and Node2Vector) in the recovery of reactions from $4$ GEMs. Reactions were removed randomly from the GEMs and treated as unobserved in the testing set.}
	\label{fig: recovery rate}
\end{figure*}

As shown in Fig.~\ref{fig: recovery rate}, CLOSEgaps exhibited an impressive recovery rate exceeding $96\%$ for all assessed GEMs. For example, with $20\%$ of reactions removed in the yeast8.5 model, CLOSEgaps effectively filled $96.82\%$ of the gaps, outperforming CHESHIRE \cite{chen2023teasing} at $94.23\%$ and Node2Vector \cite{grover2016node2vec} at $70.58\%$. In summary, CLOSEgaps displayed remarkable stability across different reaction removal percentages, indicating its ability to produce accurate results even with limited training data. These results demonstrate the practicality and potential for widespread application of CLOSEgaps in gap-filling GEMs, even with highly incomplete networks during the initial reconstruction phase.

\subsection*{Fermentation Process Improvement}\label{sec8}

\subsubsection*{Predicting Fermentation Products in Anaerobic GEMs}\label{sec13}

To gap-filling GEMs with CLOSEgaps, we employed a missing reaction inference workflow that integrates CLOSEgaps to gap-filling draft GEMs. The primary motivation behind the reconstruction of GEMs is their potential to furnish theoretical forecasts of the respective metabolic phenotypes \cite{bernstein2021addressing, ding2020novopathfinder}. Since CLOSEgaps is a deep learning-based model, which is not limited by the availability of high-quality metabolic pathways and reaction measurements \cite{zimmermann2021gapseq}.

CLOSEgaps was utilized to identify and fill metabolic gaps between the silico phenotypes and available experimental data. This was achieved through the simulation of anaerobic growth conditions and the capturing of false-positive phenotypes. To fill these gaps, we targeted reactions that were crucial for growth in the draft GEM with specific phenotypes, but are currently unexplored. Thus, we conducted biologically meaningful experiments via metabolic fermentation products for $24$ draft GEMs grown in anaerobic conditions. The draft GEMs were reconstructed using a current automatic reconstruction pipeline CarveMe \cite{machado2018fast} (\textit{SI Appendix}, Table~S3).

CLOSEgaps is trained on the entire reaction set of each draft GEM from the $24$ bacteria  \cite{zimmermann2021gapseq} with negative reactions at a $1:1$ ratio of positive reactions. Also, the candidate reaction pool includes $11,893$ reactions derived from the BiGG database. It is noteworthy that the SMILES information for these $24$ GEMs was excluded from this experiment. By adding $200$ reactions to each draft GEM resulting in the gap-filled GEM (selecting $50$, $100$, $150$, $250$, and $300$ reactions have been included in the \textit{SI Appendix}, see Fig.~S3-S8). We compared four different groups of models: the draft GEMs reconstructed from CarveMe \cite{machado2018fast}, gap-filled GEMs by adding the top $200$ reactions predicted by CLOSEgaps, CHESHIRE, and GraphSAGE from the candidate reaction pool.

\begin{figure*}[htpb]
	\centering
	\includegraphics[width=17.5cm]{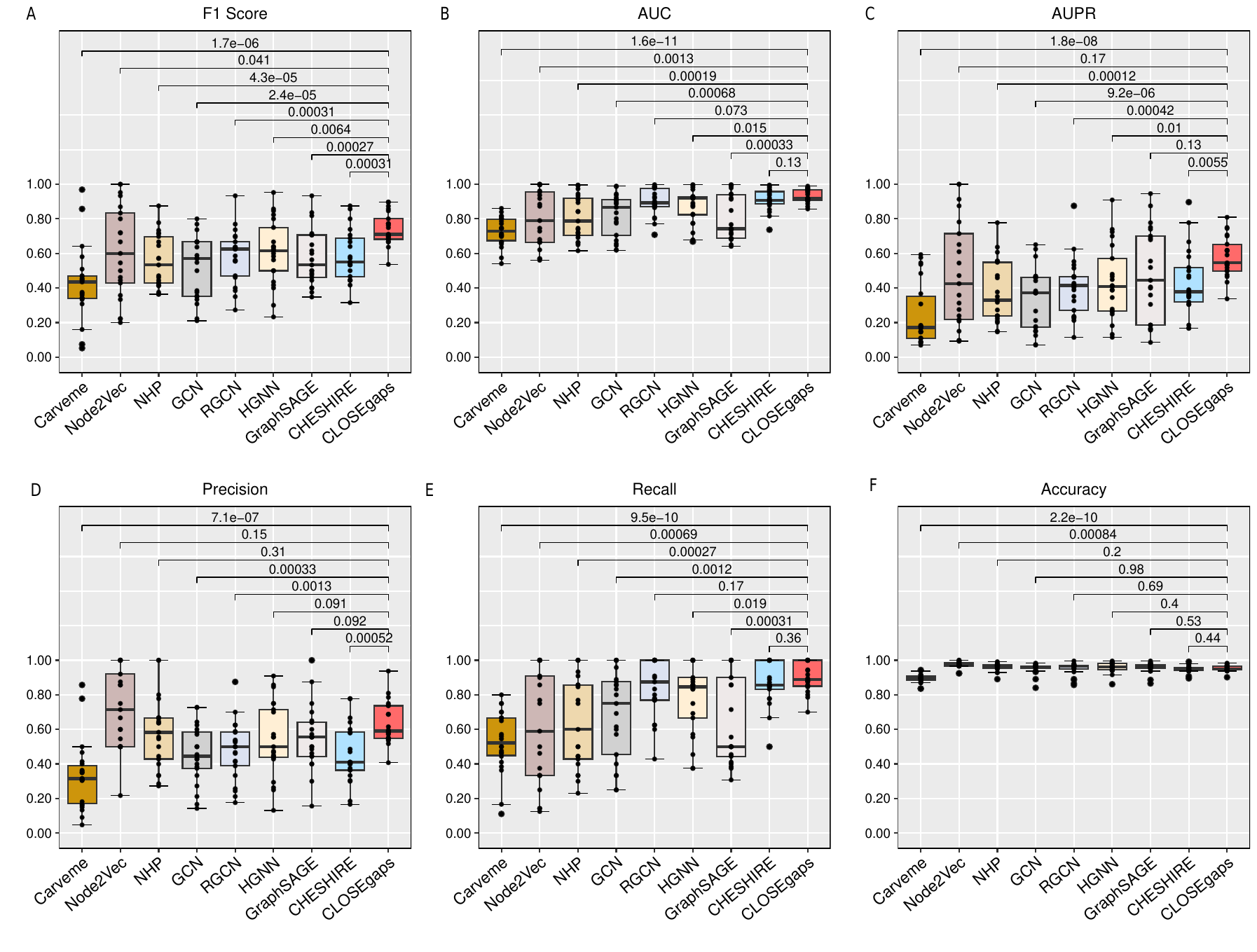}
	\caption{\textbf{Performance Comparison of Fermentation Product Predictions in Gap-Filled Metabolic Networks.} Boxplots of the performance metrics (F1 score, AUC, AUPR, Precision, Recall, and Accuracy) calculated on $24$ BiGG GEMs (each dot represents a GEM) for Node2Vector, GCN, RGCN, HGNN, GraphSAGE, CHESHIRE, and CLOSEgaps. “CarveMe” represents the draft models reconstructed from CarveMe. Each GEM was subsequently gap-filled with $200$ additional reactions predicted by each respective model. The median value for each metric is indicated by the central line in the boxplots. Statistical significance was assessed using a two-sided paired-sample t-test, with exact p-values reported.}
	\label{fig:gapfill metrics}
\end{figure*}

Using phenotype data obtained from experimental measurements assessing the synthesis feasibility of various products, we evaluated the predictive capabilities of four different groups of models: CarveMe, Node2Vector, GCN, RGCN, HGNN, GraphSAGE, CHESHIRE, and our model CLOSEgaps. As shown in Fig.~\ref{fig:gapfill metrics}, we significantly improved performance by integrating $200$ reactions ($50$, $100$, $150$, $250$, or $300$ reactions) predicted by CLOSEgaps into the draft models reconstructed from CarveMe \cite{machado2018fast}. CLOSEgaps improved the mean F1 score of CarveMe from $30.51\%$ to $64.38\%$ and the mean AUC score from $72.41\%$ to $92.90\%$. However, CLOSEgaps was unable to gap-fill three GEMs (\textit{Cutibacterium acnes} KPA171202, \textit{Clostridium acetobutylicum} ATCC 824, and \textit{Aminobacterium colombiense} DSM 12261) that each produced a single metabolite. Notably, CHESHIRE failed to fill five GEMs \cite{chen2023teasing}. The failure could happen because different GEMs from the same group may have similar missing parts. Thus, to fill these gaps, we might need to use reactions from organisms that are not closely related \cite{chen2023teasing}.

\begin{figure*}[htpb]
	\centering
	\includegraphics[width=18cm]{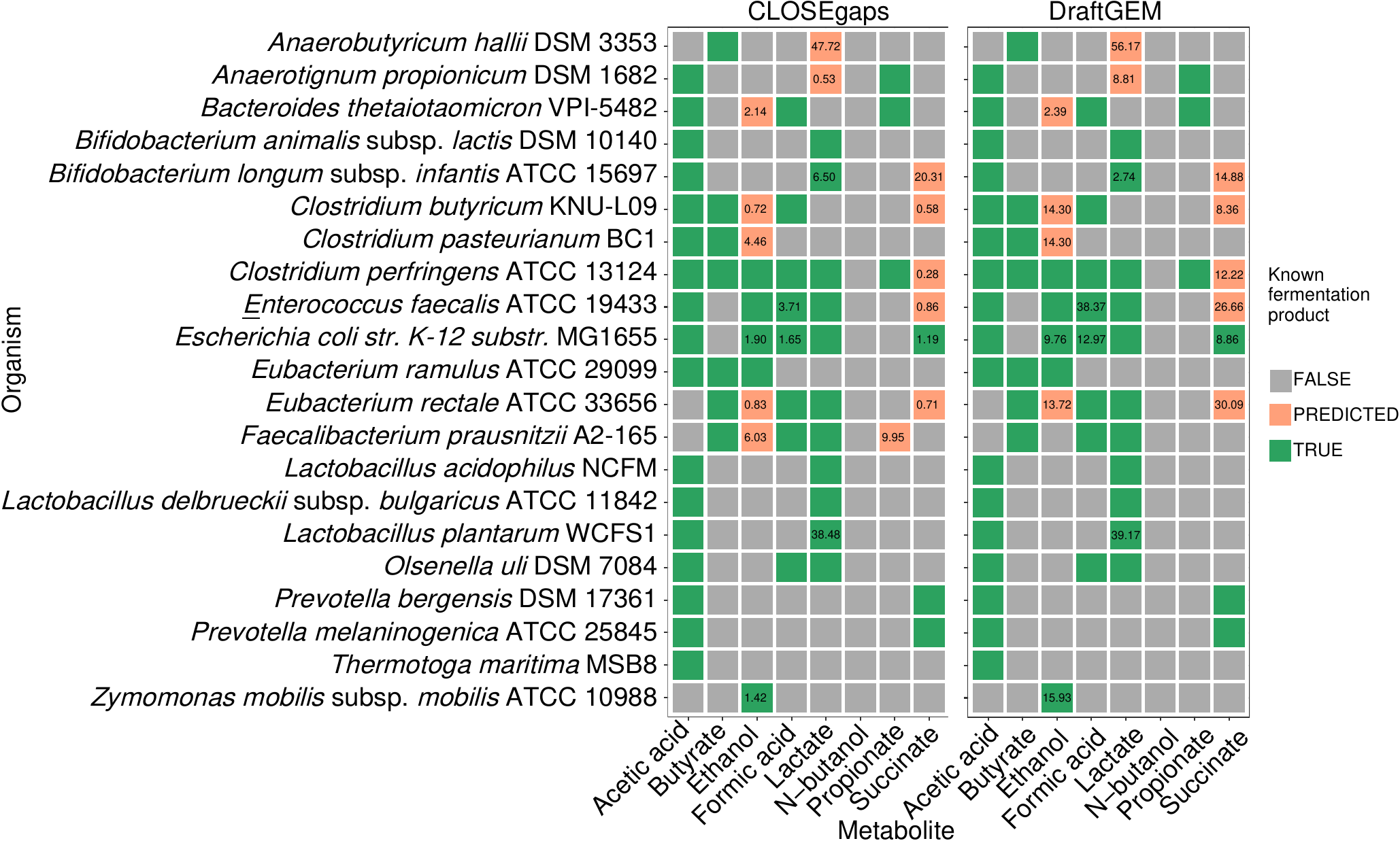}
	\caption{\textbf{Predicted anaerobic fermentation products for $24$ bacterial organisms.} A comparison of CLOSEgaps models and CarveMe (shown in the Wild-type GEM table, Excluding \textit{Cutibacterium acnes} KPA171202, \textit{Clostridium acetobutylicum} ATCC 824, and \textit{Aminobacterium colombiense} DSM 12261). Green highlights indicate known fermentation products as reported in the literature, while the orange box demonstrates the false-positive product predictions in gap-filled GEMs. The number represents the predicted metabolite production (columns) for each organism (row).}
	\label{fig:ft}
\end{figure*}

\begin{figure*}[htpb]
	\centering
        \includegraphics[width=15cm]{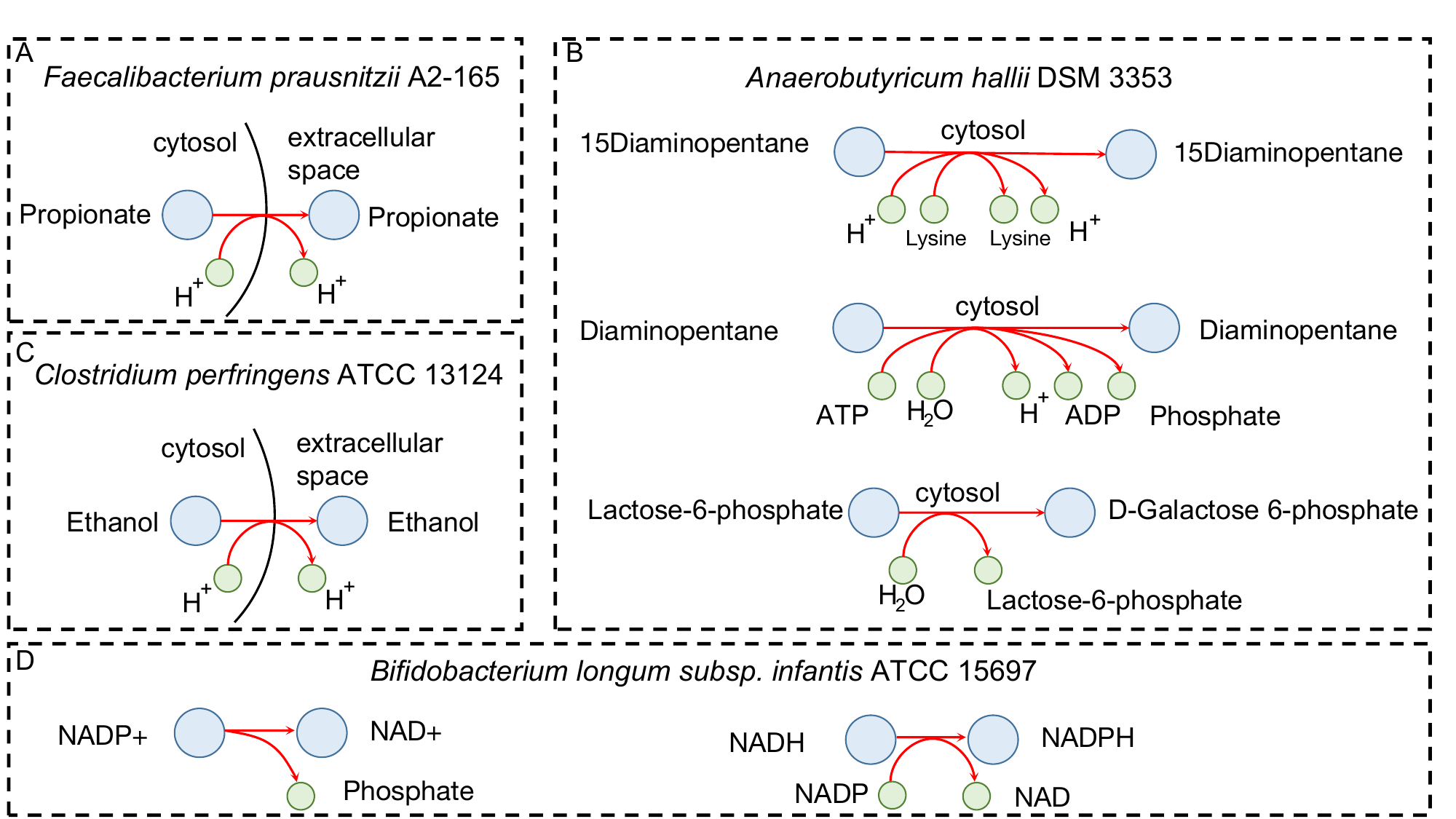}
	\caption{\textbf{(A-D) Examples of CLOSEgaps-predicted reactions that causally gap-fill the observed phenotypes.}}
	\label{fig:ft_case}
\end{figure*}

\subsubsection*{CLOSEgaps-assisted Optimization of Fermentation Pathways} \label{sec16}

Ding et al. \cite{ding2020novopathfinder} demonstrated that constructing metabolic space with novel reactions is crucial to increasing the number of value-added chemicals produced by the metabolic network. Additionally, previous research has demonstrated the potential to optimize metabolic pathways in GEMs for producing valuable fermentation products through simulation analyses (see \textit{SI Appendix}, “\textit{Assessing metabolic phenotypes and fermentation fluxes}” section) \cite{zimmermann2021gapseq}. The training process of CLOSEgaps is demonstrated in section~\ref{sec8}. CLOSEgaps ranks the hypothetical reactions within the BiGG dataset by the confidence score, which CLOSEgaps returned. Instead of adopting a fixed cutoff score, we included the $200$ reactions with the highest confidence scores. To decide whether the reaction could be rescued or not, we utilized the same strategy represented by Zimmermann et al. \cite{zimmermann2021gapseq}, that any reaction causes energy-generating cycles (EGCs) \cite{fritzemeier2017erroneous} is included (see \textit{SI Appendix}, “\textit{Ranking and rescuing hypothetical reactions}” section). This gap-filling process only rescues the reaction that these cycles could be stopped by adjusting its flux bounds. If not, skip this reaction \cite{fritzemeier2017erroneous, zimmermann2021gapseq, chen2023teasing}. Additionally, the MILP was employed to evidence the key reactions rescued for specific phenotypes were essential for growth in the draft GEM and filled previously unknown pathways in the GEM with the BiGG dataset (see \textit{SI Appendix}, “\textit{Assessing metabolic phenotypes and fermentation fluxes}” and “\textit{Identifying key reactions}” section).

We assessed the performance of CLOSEgaps-filled GEMs by comparing their production of draft GEMs constructed by Carvme. With the testing on $24$ GEMs, CLOSEgaps successfully identify and fill gaps. CLOSEgaps successfully identified $13$ false-positive phenotypes compared to available experimental data. These false positives are cases where the model predicted the production of certain metabolites that were not observed in the current dataset \cite{zimmermann2021gapseq} but are reported to be producible in the literature \cite{hao2019faecalibacterium,hoek1988physiological,jackson2012review,blacker2016investigating,fisher2015pyruvate,prasad2002crystal}. With the integration of the MILP algorithm, casually inferring the key reactions predicted by CLOSEgaps (see\textit{SI Appendix} “\textit{Identifying key reactions}” section). As shown in \mbox{Fig.~\ref{fig:ft_case}}A-D, CLOSEgaps found a metabolic network of \textit{Faecalibacterium prausnitzii} A2-165, which was expected to have a positive maximum flux for propionate but experimentally showed otherwise. Supported by previous reports \cite{hao2019faecalibacterium}. This example shows that CLOSEgaps can identify missing reactions that have consequences on distant fermentation pathways via a global and systematic effect. Gap-filling this GEM with two predicted reactions: Ethanol transport via diffusion and $H^+$/Propionate symporter (periplasm), increasing the GEMs' maximum growth rate and improving fermentation product production. As shown in Fig.~\ref{fig:ft} with the CLOSEgaps and draft GEM tables, CLOSEgaps improved the production of four metabolites (lactate, ethanol, propionate, and succinate) in two organisms, \textit{Bifidobacterium longum} subsp. \textit{infantis} ATCC 15697 and \textit{Faecalibacterium prausnitzii} A2-165. In the organism \textit{Faecalibacterium prausnitzii} A2-165, ethanol production improved to $6.03$ (mmol/gDCW/h), and propionate production improved to $9.95$ (mmol/gDCW/h). In the organism \textit{Bifidobacterium longum} subsp. \textit{infantis} ATCC 15697, succinate production increased by $5.43$ (mmol/gDCW/h), and lactate production increased by $4.03$ (mmol/gDCW/h) see \textit{SI Appendix} Fig.~S9. The reaction under consideration is catalyzed by NAD(P) transhydrogenase, an enzyme that facilitates the transfer of hydrogen ions between NADH and NADPH, thereby altering the redox states of these coenzymes \cite{hoek1988physiological,jackson2012review}. This process plays a crucial role in regulating the intracellular ratios of NAD+/NADH and NADP+/NADPH \cite{blacker2016investigating}. Although the reaction does not directly participate in the synthesis of succinate, it impacts the balance of metabolic coenzymes and the ratios of NAD+/NADH and NADP+/NADPH \cite{fisher2015pyruvate,prasad2002crystal}. This may indirectly influence various metabolic pathways within the cell, including the TCA cycle, and could potentially affect the production of succinate.

\subsection*{Evaluation of Gene Essentiality}\label{sec14}

\begin{figure*}[htpb]
	\centering
	\includegraphics[width=14cm]{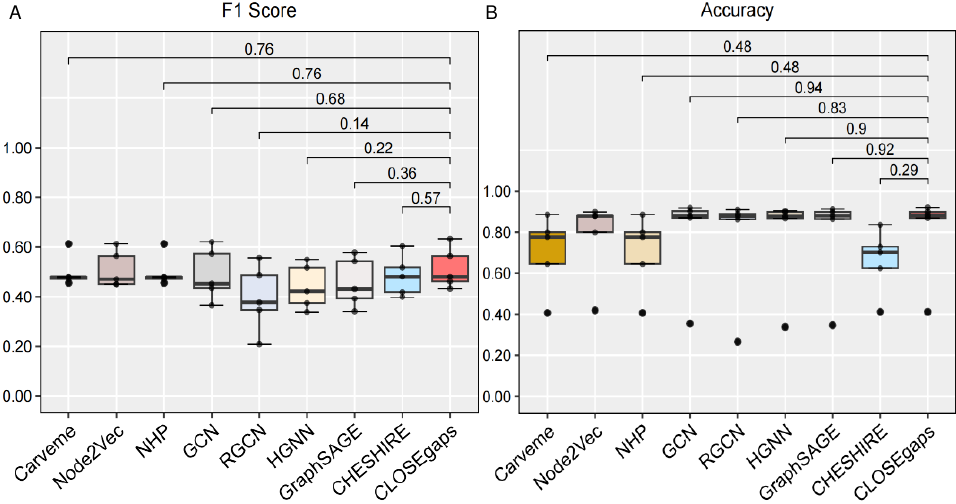}
	\caption{\textbf{Performance Comparison of Gene Essentiality.} The performance of CLOSEgaps in predicting gene essentiality was assessed using (A) F1 score and (B) Accuracy and compared against CarveMe, Node2Vector, GCN, RGCN, HGNN, GraphSAGE, CHESHIRE, and NHP across five metabolic networks.}
	  \label{gene_ess}
\end{figure*}

We compared the ability of CLOSEgaps to predict the essentiality of genes in five organisms, which were reconstructed from CarveMe (Fig.~\ref{gene_ess}). Notably, CLOSEgaps and other deep learning-based methods are gap-filling GEMs with $200$ hypothetical reactions (selecting $50$, $100$, $150$, $250$, and $300$ reactions have been included in the \textit{SI Appendix}, see Fig.~S10). Essential genes were identified through gene knockout experiments. A gene is considered essential when its removal completely stops growth \cite{zimmermann2021gapseq}. Compared to CarveMe, Node2Vector, GCN, RGCN, HGNN, GraphSAGE, and CLOSEgaps showed in all cases, a higher F1 score and accuracy in essentiality predictions. For \textit{Escherichia coli} reconstructions CLOSEgaps achieved accuracy of $92.04\%$ outperformed CarveMe with $84.55\%$ accuracy and state-of-art models CHESHIRE with $91.56\%$ accuracy. For most organisms and based on most prediction metrics, CLOSEgaps outperformed network models that were reconstructed using CHESHIRE or NHP. The results presented here consider genes as essential if the predicted growth rate of the focal gene-knockout strain was below $0.01$ h-1.

\subsection*{Pathway Visualization and Potential Reaction Heatmap}\label{sec17}

\begin{figure*}[htpb]
  \centering
      \begin{subfigure}[t]{0.91\textwidth}
        \text{A}
      \end{subfigure}
      \begin{subfigure}[t]{0.64\textwidth} \centering
      \hspace{0cm}\includegraphics[width=\linewidth, valign=t]{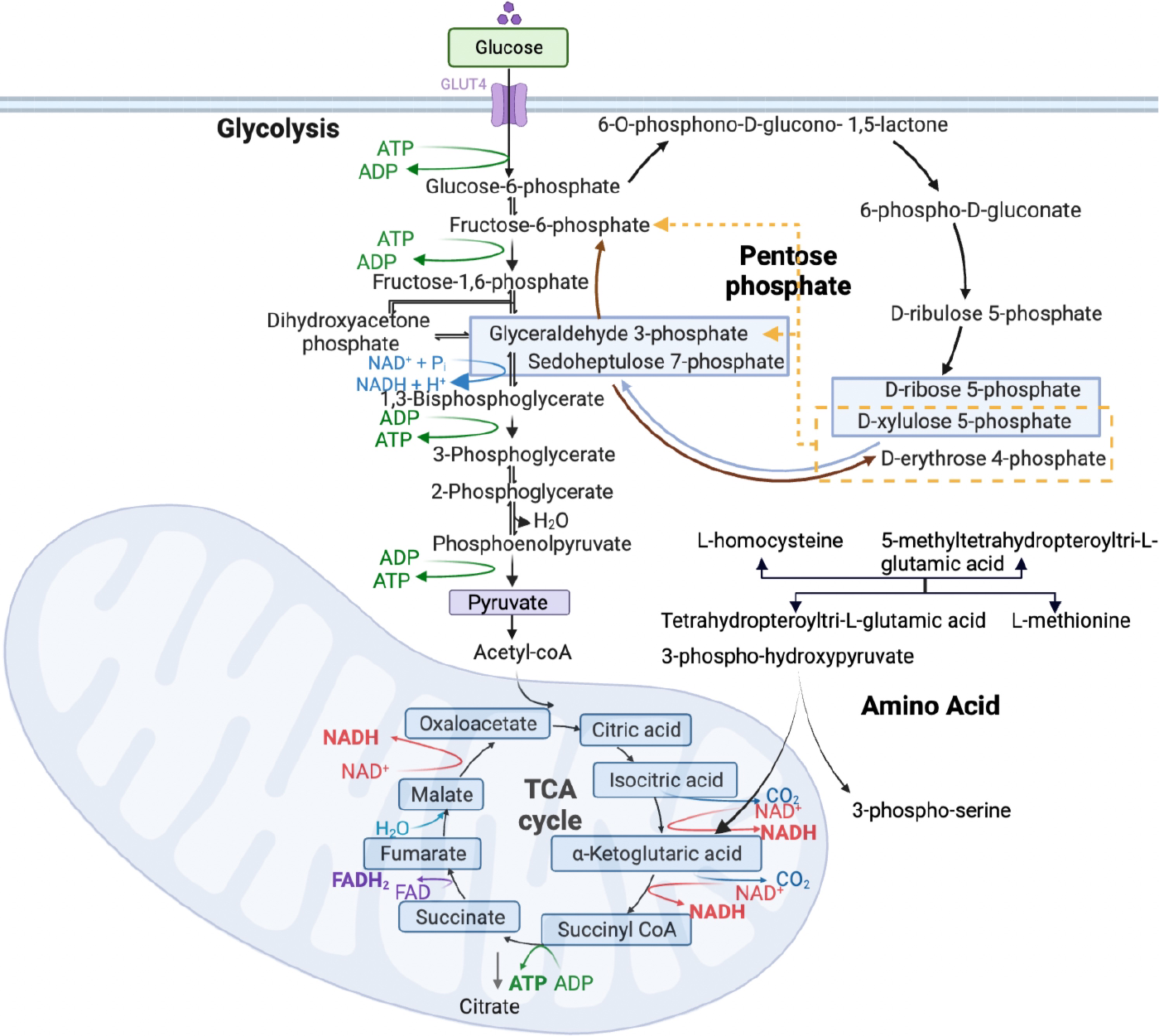}\caption{}
        \label{fig:pathwayoverview.A}
      \end{subfigure} \hfill\\
      \begin{subfigure}[t]{0.1\textwidth}
        \text{B}
      \end{subfigure}\
      \begin{subfigure}[t]{0.80\textwidth} \centering
      \hspace{0cm}\includegraphics[width=\linewidth, valign=t]{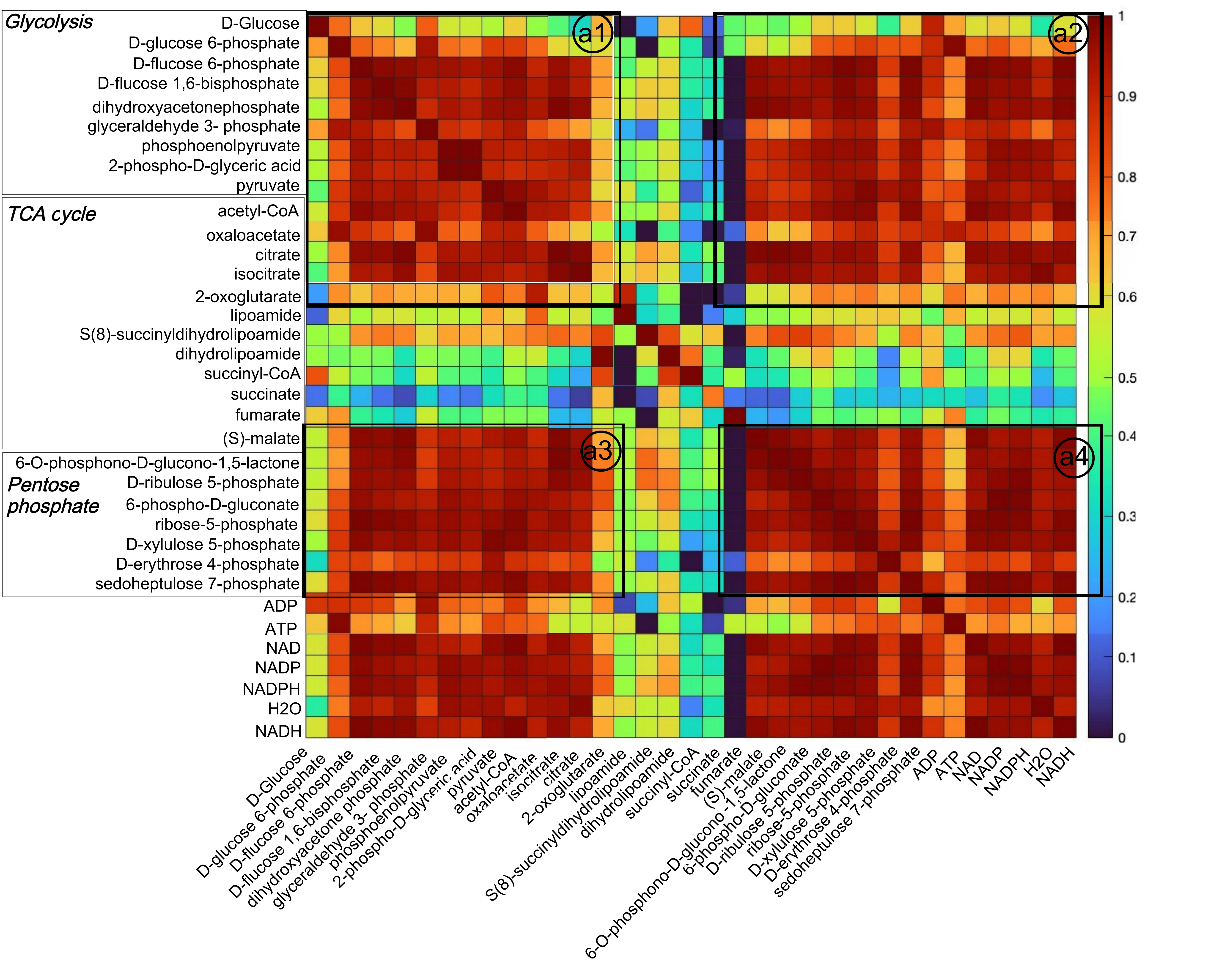}\caption{}
        \label{fig:pathwayoverview.B}
      \end{subfigure}\\
  \caption{\textbf{A} Pathway visualization of central carbon metabolism pathways and critical metabolites in \textit{S.cerevisiae}. The metabolic pathways involved glycolysis, the pentose phosphate pathway, and the tricarboxylic acid (TCA) cycle (Figure was created via Biorender.com). \textbf{B} Heatmap clustering analysis was performed to highlight potential reaction clusters. The color scale represents the strength of the relationships between metabolites, with warm colors indicating a stronger relationship.}
  \label{fig:pathwayoverview}
\end{figure*}

Moreover, CLOSEgaps offers a unique advantage in studying the central metabolic network of \textit{S.cerevisiae}. Four pathways, including glycolysis, tricarboxylic acid (TCA) cycle, pentose phosphate pathway, and part of the amino acid pathway, are visualized in Fig.~\ref{fig:pathwayoverview.A}. Meanwhile, the heatmap in Fig.~\ref{fig:pathwayoverview.B} provides a detailed visualization of the feature embeddings generated by CLOSEgaps, serving as a clear and intuitive tool for identifying potential reactions based on metabolite relationships and enhancing the interpretation of key metabolic pathways. This figure illustrates interactions between metabolites within glycolysis (a1), the TCA cycle (a2), the pentose phosphate pathway (a3), and the integrated pathways (a4) (see \textit{SI Appendix}, Fig.~S11A, B, C, and D for each reaction type). Warmer color regions highlight potential reaction clusters, indicating metabolites likely to participate in the same reaction. Briefly, Fig.~\ref{fig:pathwayoverview.A} and Fig.~\ref{fig:pathwayoverview.B}a1 illustrate the complex process of glycolysis, the warm clusters indicate strong interactions among key metabolites such as D-Glucose and D-Glucose 6-phosphate \cite{mao2023cave}. Within the TCA cycle (Fig.~\ref{fig:pathwayoverview.B}a2), CLOSEgaps effectively captures the significant reactivity of (S)-malate, while succinate and fumarate exhibit lower intensity of interaction, as reflected by its cooler. Compared to succinate and fumarate, the heightened reactivity of (S)-malate in central metabolism is attributed to its interconversion with pyruvate within the glycolytic pathway and its metabolic transformation with oxaloacetate, a pivotal precursor in the TCA cycle. In the pentose phosphate pathway (Fig.~\ref{fig:pathwayoverview.B}a3), the model accurately identifies the heightened reactivity of metabolites like 6-O-phosphono-D-glucono-1,5-lactone and D-ribulose 5-phosphate. CLOSEgaps demonstrates its capability to discern between highly reactive clusters and less interactive regions, particularly in cross-pathway interactions (Fig.~\ref{fig:pathwayoverview.B}a4), underscoring its effectiveness in modeling the complex dynamics of metabolic networks.

In addition, CLOSEgaps distinguishes itself from conventional graph embedding methods by being able to learn a high-order, relation-aware embedding for link prediction. We provide a comprehensive visual representation of interactions among various metabolites across key metabolic pathways, including glycolysis, the pentose phosphate pathway, and the TCA cycle (see \textit{SI Appendix}, Fig.~S11). CLOSEgaps effectively groups related metabolites within specific pathways, as demonstrated by blocks of similar colors, reflecting biochemical connectivity. Such distinctive clustering and the clear demarcations between different metabolic pathways underscore the accuracy of CLOSEgaps in metabolic profiling and its utility in experimental planning. Additionally, variations in color intensity across the heatmap highlight the sensitivity and precision of CLOSEgaps, enabling a fine-grained analysis of metabolic interactions. The quantitative gradient scale enhances the utility by providing a measurable method to assess and compare interaction strengths. 

To better illustrate the difference between CLOSEgaps and other baseline methods, such as Node2Vector and GCN, we present a visual demonstration of the learned reaction embedding through the t-SNE tool \cite{van2008visualizing} (see \textit{SI Appendix}, Fig.~S12). Overall, the ability of CLOSEgaps to dissect and display detailed metabolic interactions enhances understanding of complex metabolic networks.

\section{Discussion}
In this study, we introduce a new and innovative metabolic network reconstruction workflow and machine learning model, NICEgame and CLOSEgaps, which is aimed at predicting missing reactions. CLOSEgaps is a fully data-driven model that is built on the foundation of curated GEMs and hypothetical reaction data. The comprehensive evaluations of both internal and external test sets reveal that our method effectively curates GEMs, leading to improved predictions of missing metabolic reactions and functions that can later be verified through experimentation. Our approach demonstrated a high mean recovery rate of $95.34\%$ when benchmarked on the yeast8.5 model using artificially introduced gaps. Additionally, by integrating mixed-integer linear programming, we further benchmarked CLOSEgaps using fermentation data and improved the prediction of fermentation products for $24$ bacterial organisms, achieving a mean F1 score of $74.24\%$ based on wild-type GEMs reconstructed from CarveMe \cite{machado2018fast}. The use of CLOSEgaps resulted in improved production of ethanol and propionate in \textit{Faecalibacterium prausnitzii} A2-165, and of succinate and lactate in \textit{Bifidobacterium longum} subsp. infantis ATCC 15697. These results highlight the potential of CLOSEgaps as a valuable tool in optimizing fermentation pathways and metabolic network reconstruction.

CLOSEgaps address a crucial need for efficient GEM curation to enhance \textit{in silico} predictions of missing metabolic reactions and functions. There is room for future improvement through the provision of additional information, advanced enzyme prediction \cite{levin2022merging, li2022deep}, and design tools \cite{jumper2021highly, huang2016coming, huang2022backbone, li2023gotenzymes} to aid in experimental analysis. This work has significant implications for the study of metabolic networks. CLOSEgaps can be applied to any existing GEM, advancing the fields of biotechnology and biomedicine. The key reactions predicted by CLOSEgaps can serve as a valuable resource for identifying new ways to improve strain performance, such as increasing biomass or product yield. CLOSEgaps holds tremendous potential for identifying metabolic network gaps, reconstructing metabolic networks, and rational design.

\section{Materials and methods}

\subsection*{Universal Bacterial Models and Biochemistry database} \label{sec2}
The workflow of CLOSEgaps includes four databases (see Fig.~\ref{fig1}A): GEM of the organism, Metabolites reference SMILES dataset, Metabolites database, and BiGG candidate reactions database. Specifically, the draft GEM models were downloaded from the BiGG database (http://BiGG.ucsd.edu) in August 2022 \cite{norsigian2020bigg}. To represent metabolites and reactions in the SMILES string format, each GEM includes a crafting and cleaning process with our manually generated metabolites SMILES database (see Fig.~S1, \textit{SI Appendix}). In addition, to generate the negative samples for model training, we curated $44,359$ metabolites with valid SMILES sequence from the ChEBI dataset (https://www.ebi.ac.uk) as metabolites database (see Data collection and preprocessing section, \textit{SI Appendix}). 

To assess the generalization ability of CLOSEgaps, the chemical reactions are also included, sourced from the full USPTO chemical reaction dataset consisting of $1.8$ million reaction represented in SMILES notation, recorded from 1976 to 2016 \cite{lowe2012extraction}. To assess how the prediction accuracy is influenced by dataset size, we construct two chemical reaction datasets for our experiments by randomly selecting $3,000$ and $8,000$ reactions from Lowe, D.M. \cite{lowe2012extraction}. The statistics of datasets are summarised in Table~\ref{dataset_statistics}. Moreover, $11,893$ reactions (hypothetical reactions) were downloaded from the BiGG database, which was collected from $79$ metabolic networks of various organisms, forming a candidate reaction pool for our experiments. Notably, reactions classified as biomass, exchange, demand, sink, or already present in the network were systematically removed.

\subsection*{Problem Description}

CLOSEgaps primarily utilizes deep learning-based algorithms to represent GEMs as hypergraphs and gap-filling GEMs through hypergraph learning. This process involves predicting missing reactions through hyperlink prediction. Gap-filling via CLOSEgaps relies on topology information, which aids in identifying gaps and uncovering the missing reactions even for highly incomplete GEMs. To describe the problem with a given GEM, we define the hypergraph as $\mathcal{H}=\{\mathcal{V},\mathcal{E}\}$, where $\mathcal{V}=\{v_{1},v_{2},...,v_{n}\}$ is the node (metabolite) set and hyperedges $\mathcal{E}=\{e_{1},e_{2},...,e_{m}\}$, $e_{\epsilon} \subseteq \mathcal{V}, \epsilon=1,2,...,m$ for the metabolic network. In general, the hypergraph $\mathcal{H}$ can be represented by its incidence matrix $\mathbf{H}_p \in \mathbb{R}^{n \times m}$, each row corresponds to a metabolite and each column to a reaction. An entry of $1$ indicates the participation of a metabolite in a reaction, while $0$ indicates no participation. This incidence matrix is derived by converting non-zero entries in the stoichiometric matrix into binary values, thus simplifying the representation of the GEM into a hypergraph structure \cite{chen2023teasing, yadati2020nhp}. The problem of hyperlink prediction in the incomplete undirected hypergraph $\mathcal{H}$ involves predicting missing hyperlinks from $\mathcal{\bar{E}}=2^{V}-E$ based on the current set of observed hyperlinks $\mathcal{E}$.

We compared CLOSEgaps with the state-of-the-art machine learning methods CHESHIRE \cite{chen2023teasing}, GraphSAGE \cite{schlichtkrull2018modeling}, and NHP \cite{yadati2020nhp} as they have been demonstrated to display superior performances for link prediction. We also include relatively simple learning architecture models, such as GCN and Node2Vector \cite{grover2016node2vec,kipf2016semi} to demonstrate the superior or hypergraph learning. In addition, to demonstrate the superiority of the multi-layer and multi-head mechanism, we compared CLOSEgaps with the relational graph-based model RGCN \cite{schlichtkrull2018modeling} and the basic hypergraph learning model HGNN \cite{feng2019hypergraph}. These models use different graph embedding methods that generate node features and mean pooling to generate metabolite and reaction features.

\subsection*{CLOSEgaps Model}

As shown in Fig.~\ref{fig1}a, the architecture of CLOSEgaps consists of five modules: database construction, negative reaction sampling, feature initialization, feature refinement, and ranking hypothetical reactions. Furthermore, a generalized workflow for automated metabolic network reconstruction comprises three stages (Fig.~\ref{fig1}d). The first stage maps metabolites to SMILES \cite{weininger1988smiles} using the public database of Biochemistry (see \textit{SI Appendix,} Fig.~S1). The second stage uses CLOSEgaps to rank and add top $N$ reactions to the draft GEMs to create gap-filled GEMs ($N$ reactions with the highest confidence scores). Specifically, CLOSEgaps is trained on the full reaction set of the draft GEM, while candidate reactions from a reaction pool (e.g., the BiGG database) are ranked based on confidence scores. The third stage applies flux simulation to predict metabolic phenotypes. CLOSEgaps tries to match the predictions of the draft model with the observed phenotype by adding reactions. Suppose there is a discrepancy between the gap-filled and draft GEMs. In that case, it indicates unexplored pathways, which are addressed in the workflow using Mixed Integer Linear Programming (MILP) \cite{gu2019current, domenzain2022reconstruction} to infer the reactions causing the gaps.

To improve the ability of the deep-learning model to predict missing reactions, we incorporated negative sampling by generating non-existent reactions (see Fig.~\ref{fig1}b and \textit{SI Appendix}, Negative sampling strategies) \cite{mikolov2013distributed, zhang2018beyond}. This involves substituting percentages of metabolites from a positive hyperlink with those from the ChEBI database, expanding our hypergraph to include these negative samples, making $\mathbf{H}=[\mathbf{H}{p}|\mathbf{H}{n}] \in \mathbb{R}^{n \times 2m}$. The positive and negative reactions for each GEM were combined and randomly and randomly split into training, validation, and testing sets over $10$ Monte Carlo runs ($60\%$ as the training set, $20\%$ for validation, and $20\%$ for testing. The process was repeated $10$ times).

To initialize the hypernode feature (see Fig.~\ref{fig1}c) represented as $\mathbf{X}^{(0)}$. The hyperedge feature denoted $\mathbf{X}^{e}$, we used a fully connected layer based on three matrices: the GEM hypergraph matrix $\mathbf{H}_p$, the initial hypergraph matrix $\mathbf{H}$ combined with negative reactions, and the metabolite similarity matrix $\mathbf{S}$. The similarity matrix $\mathbf{S} \in \mathbb{R}^{n \times n}$ was constructed using the Tanimoto coefficient with SMILES sequence \cite{ma2022hypergraph}. The initial hypernode feature $\mathbf{X}^{(0)}$ is defined in Equation~(\ref{fn}), the hyperedge feature $\mathbf{X}^{e}$ (Equation~(\ref{fe})), is embedded using a fully connected layer that integrates with the transpose of the incidence matrix $\mathbf{H}$, represented as $\mathbf{H}^{T}$:

\begin{subequations}\label{l1} 
\begin{align}
&\mathbf{X}^{(0)} = \text{Cat}(\text{Linear}(\mathbf{H}_p),\text{Linear} (\mathbf{S})),\label{fn}\\
&\mathbf{X}^{e} = \text{Linear}(\mathbf{H}^{T}),
\label{fe}
\end{align}
\end{subequations} 

where the $\text{Linear}(\cdot)$ represents the application of a fully connected layer, whereas $\text{Cat}(\cdot)$ represents the concatenation operation. 

For the feature refinement process, inspired by Feng et al. \cite{feng2019hypergraph}, we enhanced the framework with a multi-channel hypergraph convolution network and a multi-head attention module (see Fig.~\ref{fig1}c.1 and c.2) to capture high-order interactions. Following a previous study, Bai et al. \cite{bai2021hypergraph} simplify this by focusing on interactions between points linked by the same hyperedge and prioritizing hyperedges with higher weights. For the given hypergraph, each hyperedge $e_{\epsilon} \in \mathcal{E}$ is associated with a positive weight $W_{\epsilon\epsilon}$.

Subsequently, one step of hypergraph convolution is defined in Equation~(\ref{conv}). In Equation~(\ref{X1}), $x_{i}^{(l)}$ represents the embedding of the $i$th vertex at the $(l)$th layer, which is refined to Equation~(\ref{x2}), as the matrix $\mathbf{P} \in \mathbb{R}^{F^{(l)} \times F^{(l+1)}}$ is the weight matrix between the $(l)$th and $(l+1)$th layers. Thus, hypergraph convolution is defined in Equation~(\ref{x3}):

\begin{subequations}\label{conv}
\begin{align} 
&x_{i}^{(l+1)} = \sigma(\sum\nolimits_{j=1}^{n}\sum\nolimits_{\epsilon =1}^{2m}H_{i \epsilon}H_{j \epsilon}W_{\epsilon \epsilon} x_{j}^{(l)}\mathbf{P}),\label{X1}\\
&\mathbf{X}^{(l+1)} = \sigma (\mathbf{HW}\mathbf{H}^{T}\mathbf{X}^{(l)}\mathbf{P}), \label{x2} \\
&\mathbf{X}^{(l+1)} = \sigma (\mathbf{D}^{-1}\mathbf{HWB}^{-1}\mathbf{H}^T\mathbf{X}^{(l)}\mathbf{P}), \label{x3}
\end{align}
\end{subequations}

where $\mathbf{X}^{(l)}$ and $\mathbf{X}^{(l+1)}$ are the input of the $(l)$th and $(l+1)$th layer, respectively. And the the node degree is defined as $\mathbf{D}_{ii} = \sum\nolimits_{\epsilon =1}^{2m}W_{\epsilon \epsilon}\mathbf{H}_{i\epsilon}$, the hyperedge degree is $\mathbf{B}_{\epsilon \epsilon} = \sum\nolimits_{i=1}^{n}\mathbf{H}_{i \epsilon}$. Note that $\mathbf{D} \in \mathbb{R}^{n \times n}$ and $\mathbf{B} \in \mathbb{R}^{2m \times 2m}$ are both diagonal matrices. And $2m$ represents the number of hyperedges and $n$ denotes the number of nodes. The function $\sigma \left ( \cdot \right )$ denotes a non-linear activation function, such as ReLU. Attention scores between nodes and hyperedges are computed using a similarity function as shown in Equation~(\ref{a1}):

\begin{equation}
H_{i\epsilon} = \frac{\text{exp}\left ( \sigma \left ( \text{sim}\left ( x_{i}^{(l)}\mathbf{P},{x^{e}_{\epsilon}}^{(l)}\mathbf{P} \right ) \right ) \right )}{\sum_{k \in \mathcal{N}_{i}}\text{exp}\left ( \sigma \left ( \text{sim}\left ( x_{i}^{(l)}\mathbf{P}, {x^{e}_{k}}^{(l)}\mathbf{P} \right ) \right ) \right )},
\label{a1}
\end{equation}

where $\mathcal{N}_{i}$ is the neighborhood set of $v_i$. A similarity function denoted as $\text{sim} \left ( \cdot  \right )$ is employed to compute the pairwise similarity between each pair of nodes.

\subsection*{Ranking and rescuing hypothetical reactions}\label{sec7}

Predicting the absence or presence of hypothetical reactions within the GEM was structured as a binary classification task (see Fig~\ref{fig1}c.3). Each feature vector of the hyperlink was fed into a softmax function to predict a probability distribution across two classes: existence or non-existence. Specifically, to incorporate metabolite features into a hyperlink-level representation, we utilize matrix multiplication with the refined hypernode feature $\mathbf{X}^{(L)}$, where $L$ is the number of hypergraph convolution layers and the transpose of hyperedge feature matrix $\mathbf{H}^{T}$, represented as Equation~(\ref{softmax}):

\begin{equation}
    \mathbf{Y} = \text{Softmax}(\text{Linear}(\mathbf{H}^{T}\mathbf{X}^{(L)})),\label{softmax}
\end{equation}

where $Y=\{y_1,y_2,...,y_{2m}\}$ represents the prediction score for each hyperlink. $\text{Linear}(\cdot)$ and $\text{Softmax}(\cdot)$ present the fully connected layer and softmax function, respectively.

To address the deficiencies in a draft GEM, we implemented a missing reaction inference workflow using a pool of hypothetical reactions from the BiGG database to generate a gap-filled GEM. All draft GEMs of those organisms were reconstructed using a recent automatic reconstruction pipeline, CarveMe \cite{zimmermann2021gapseq}. To generate the gap-filled GEMs, only growth phenotypes were used with a compiled dataset including fermentation profiles of eight metabolites from twenty-four bacterial organisms grown under anaerobic conditions. 

Firstly, the collected BiGG reaction pool includes $11,893$ reactions. CLOSEgaps was deployed to rank each reaction, providing a confidence score that quantified the probability of a particular reaction requiring rescue for the currently tested draft GEM. Particularly, a very stringent threshold of $0.99999$ was chosen to rank reactions. Secondly, instead of using a fixed cutoff score, gap-filling GEM iteratively adds the top $200$ reactions, causing energy-generating cycles (EGCs) with the highest confidence scores. We employed the method by Fritzemeier et al., \cite{fritzemeier2017erroneous} to detect EGCs in gap-filled GEMs derived from wild-type GEMs; the reaction was included if EGCs can be eliminated by changing its flux bounds and otherwise skipped \cite{fritzemeier2017erroneous, zimmermann2021gapseq, chen2023teasing}. This approach generates $15$ energy dissipation reactions for ATP, CTP, GTP, and other energy metabolites. The presence of a non-zero flux in a dissipation reaction signifies an EGC. In the case of reversible reactions, we constrained their flux, whereas irreversible reactions were skipped. Additionally, reactions involving oxygen were omitted due to anaerobic growth conditions. This process was repeated until all $200$ reactions had been added.

\subsection{Training Algorithm} \label{sec8}
We leverage the exceptional efficiency of the Adam optimization algorithm \cite{jais2019adam} to train CLOSEgaps with the following cross-entropy loss function as Equation~(\ref{loss}):
\begin{equation}
\text{Loss} = \frac{1}{2m}(\sum_{e_i \in \mathcal{E}_p}\log(y_i)+\sum_{e_i \in \mathcal{E}_n}\log(1-y_i)),
\label{loss}
\end{equation}

where $\varepsilon_{p} $ is the set of positive hyperlinks, $\varepsilon_{n} $ is the set of negative hyperlinks. In the training phase, CLOSEgaps learns the weights of the deep neural network by minimizing the loss function via maximizing the scores for positive hyperlinks. Throughout the evaluation phase, CLOSEgaps utilizes the learned weights to compute a probability score for an unseen hyperlink originating from either a testing set or a comprehensive candidates reaction pool dataset.

\subsection*{Data, Materials, and Software Availability}
All study data are included in the article and/or supporting information. The raw data is collected from ChEBI (https://www.ebi.ac.uk/), and BiGG (http://bigg.ucsd.edu/). The code and model have been provided in Github [\url{https://github.com/guofei-tju/CLOSEgaps}]

\showmatmethods{} 

\acknow{This study was supported by the grant from National Natural Science Foundation of China (NSFC 62322215, 62172296, 62172076), Excellent Young Scientists Fund in Hunan Province (2022JJ20077), and Shenzhen Science and Technology Program (No.KQTD20200820113106007). Also, this work was supported in part by the High Performance Computing Center of Central South University.}

\showacknow{}

\end{document}